\documentclass[twocolumn,superscriptaddress,aps,showpacs,amsmath,footinbib,bibnotes]{revtex4-1}
\pdfoutput=1
\usepackage{graphicx}
\usepackage{amssymb,amsmath}
\usepackage{textcomp} 
\usepackage[bold]{hhtensor}
\usepackage{natbib}
\usepackage{lipsum}
\usepackage{bm}
\usepackage{hyperref}
\usepackage{multirow}
\usepackage{float}
\usepackage{amsmath}
\usepackage{amssymb}
\usepackage{stmaryrd}
\usepackage{mathrsfs}
\usepackage[dvipsnames]{xcolor}
\hypersetup{
	colorlinks,
	linkcolor={red!80!black},
	citecolor={Blue},
	urlcolor={MidnightBlue}
}

\begin{document}

\author{Boris Desiatov}
\affiliation{John A. Paulson School of Engineering and Applied Sciences, Harvard University, Cambridge, Massachusetts 02138, USA}
\author{Marko Lon{\^{c}}ar}\email{loncar@seas.harvard.edu}
\affiliation{John A. Paulson School of Engineering and Applied Sciences, Harvard University, Cambridge, Massachusetts 02138, USA}

\date{\today}
\title{Silicon photodetector for integrated lithium niobate photonics }

\begin{abstract}
We demonstrate integration of an amorphous silicon photodetector with thin film lithium niobate photonic platform operating in the visible wavelength range. We present the details of the design, fabrication, integration and experimental characterization of this metal-semiconductor-metal photodetector that features responsivity of 22 mA/W to 37mA/W over the wide optical bandwidth spanning 635 nm to 850 nm wavelengths range.
\end{abstract}

\maketitle

Integration of various photonic components on a single chip, including light sources and detectors, is a critical route towards realization of dense photonic integrated circuits (PIC) [1]. These are of interest not only for traditional applications in data- and tele-communications, but also for applications in imaging, metrology, bio-sensing, nanomedicine and quantum optics that typically require operation in the visible wavelength range[2,3]. For decades, lithium niobate has been considered to be an optimum optical material due to it large second order ($\chi$-2) electro-optic coefficient and excellent wideband optical transparency (400 nm – 4,000 nm). However, traditional LN photonic structures, created by ion exchange or metal in-diffusion, suffer from the low refractive index contrast, resulting in a large cross-section of the photonic structures thus making the dense integration difficult. Recently, thin film lithium niobate on insulator (LNOI) substrates[]4] have become commercially available, which combined with advances in nanofabrication has enabled realization of ultra-low loss waveguides and high-performance electro-optical (EO) devices both at telecom[5,6] and visible wavelengths[7].  Integration of photodetectors (and eventually laser sources) with LNOI photonic platform is an important, and currently missing, step that could lead to implementation of complex functionalities[8]  using this emerging PIC platform. Indeed, opportunities offered by integration of detectors with lithium niobate have been explored before[9,10] albeit using traditional, bulk crystal based, in-diffused LN waveguides with low optical confinement and large mode size. This results, among other things, in large detector area thus limiting the response time of the photodetector. In this paper, we demonstrate monolithic integration of broadband Metal-Semiconductor-Metal (MSM) photodetector for visible wavelengths in thin film LNOI photonic platform.   

\begin{figure}
	\centering
	\includegraphics[angle=0,width=0.5\textwidth]{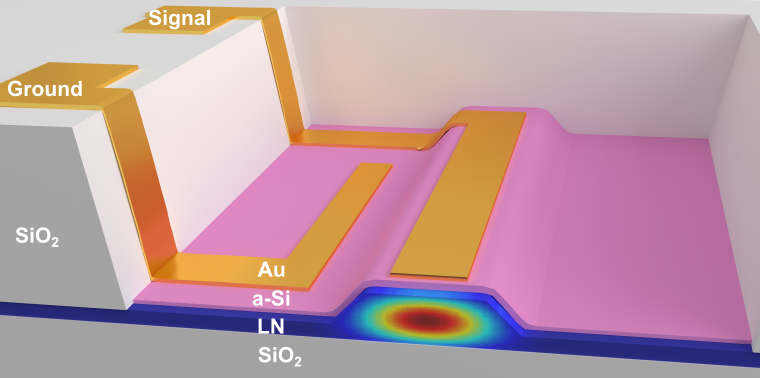}
	\caption{\label{fig1} An illustration of integrated photodetector device consisting of the LN waveguide with cross section 800x300nm, a-Si absorption layer and gold contacts.
}

\end{figure}

Over the past decade, significant advances have been made along the lines of integration of semiconductor photodetectors with photonics platforms operating in the visible wavelength range, motivated by envisioned applications in bio-sensing and imaging[11$-$15].Compared to PN or PIN junction-based photodetectors used in these experiments, MSM[16,17] junction-based photodetector described in this work is relatively simple to implement since it does not require dopant implantation[16,17]. Still, MSM photodetectors  can have a high responsivity, low capacitance, low dark current and high operation speed comparable to, and even better than, modern PN or PIN photodetectors[16,17]. Previously, such type of a-Si photodetector at visible wavelength was demonstrated[18] in sputtered glass waveguides back in 1988. A schematic of our integrated MSM photodetector is shown in Fig.1. It consists of thin layer of amorphous silicon (a-Si) deposited on top of LN waveguide, with a pair of gold electrodes on top of it. Photons at visible wavelength, propagating down the LN waveguide, are absorbed in the a-Si layer, and generated electron-hole pairs are separated by voltage applied across the MSM junction giving rise to a photo-current.

The most important parameter which directly affects the performance of the proposed photodetector is a thickness of a-Si absorption layer. A very thick a-Si layer ensures that all the optical power propagating in the waveguide is effectively absorbed, but transport and collection of generated photocarriers becomes more challenging due to recombination processes. On the other hand, if a very thin layer of a-Si is used, optical field can have significant overlap with metal electrodes which may result in unwanted ohmic losses. Though, it is important to note that interaction of optical energy with a metal contact can also contribute to the generation of electron-hole pairs due to internal photoemission effect[19,20] (generation of hot carriers).

Our rib LN waveguide has a dimensions of 800x300nm and can support both fundamental transverse electric (TE) and transverse magnetic (TM) modes as shown in Fig.2(a) and Fig.2(c) respectively. The corresponding TE and TM modes of photodetector structure are shown in Fig.2(b) and Fig.2(d) respectively.  In order to find an optimal thickness of absorption a-Si layer we have performed a detailed study of optical absorption length and parasitic Ohmic loss in the metal as a function of the a-Si layer thickness, for both polarizations of interest. Using a 3-dimensional (3D) FDTD simulation (Lumerical. Inc.) we calculated the total absorption length (defined as distance where the incident power drops to 1/e) and a contribution of optical absorption in the metal (due to ohmic losses) for several a-Si layer thicknesses, for both fundamental TE and TM modes at wavelength of 635nm. The permittivity of gold material is based on data from Johnson and Christy[21]. The simulated results are summarized in Fig.2(e)\  and Fig.2(f) respectively. For a thickness of a-Si below 50nm, both TE and TM modes have a very high metal loss (more than 50$\%$ ) making such thin semiconductor layer ineffective for realization of MSM detector. For the TE mode and moderately thick semiconductor layer (above 100nm) the optical absorption in the metal region can be neglected (less that 5$\%$ ) and the total absorption length was found to be below 10$ \mu $m. For the TM mode the loss in the metal region of the detector was found to be relatively high ($ \sim $  10$\%$ ) even for a relatively thick a-Si layer of 300nm. Such big difference in the metal-induced optical losses in the case of TE and TM modes can be explained by different boundary condition at the semiconductor – metal interface. We choose to work with 100nm thick a-Si layer, and have calculated the coupling efficiency between the low-loss LN waveguide and a photodetector to be 89$\%$. A cross section sideview of optical simulations for the optimal parameters is shown in Fig.2(g). The reflected and scattered amount of optical power was found to be 7$\%$  and 4$\%$ , respectively. These numbers can be further decreased by introducing an adiabatic tapering transition section between LN waveguide and the photodetector. The absorption length of the photodetector was found to be 6 $ \mu $m.

The electrical bandwidth of MSM photodetector can be limited by the capacitance of the electrodes :\   \( C= \varepsilon _{0} \varepsilon _{aSi}\frac{Lt}{d} \)  or by carrier collection time [22]  \(  \tau_{collect}=\frac{d}{v_{sat \_ aSi}}  \) . Here,  \(  \varepsilon _{0} \)  is the vacuum dielectric constant, the  \(   \varepsilon _{aSi} \)  relative dielectric constant,  \( v_{sat \_ aSi}  \) the a-Si carrier saturation velocity, \(  L \)  the photodetector length,  \( d \)  the distance between the electrodes, and  \( t \)  the photodetector thickness. For a 1 $ \mu $m distance between the electrodes, the capacitance of the photodetector was calculated to be 0.5fF, resulting in the RC time constant of 25ps (for 50$ \Omega $  load). The saturation velocity of a-Si layer strongly depends on deposition/annealing parameters used during the fabrication process. Here, we assume carrier collection time of our detector to be on the order of nanosecond, based on  previously reported value for saturation drift velocity of 8.5x10{5}cm/V[23] (and taking into account the detector geometry).This value of carrier collection time in our a-Si MSM photodetectors is similar to the values previously reported[24].

\begin{figure}
	\centering
	\includegraphics[angle=0,width=0.5\textwidth]{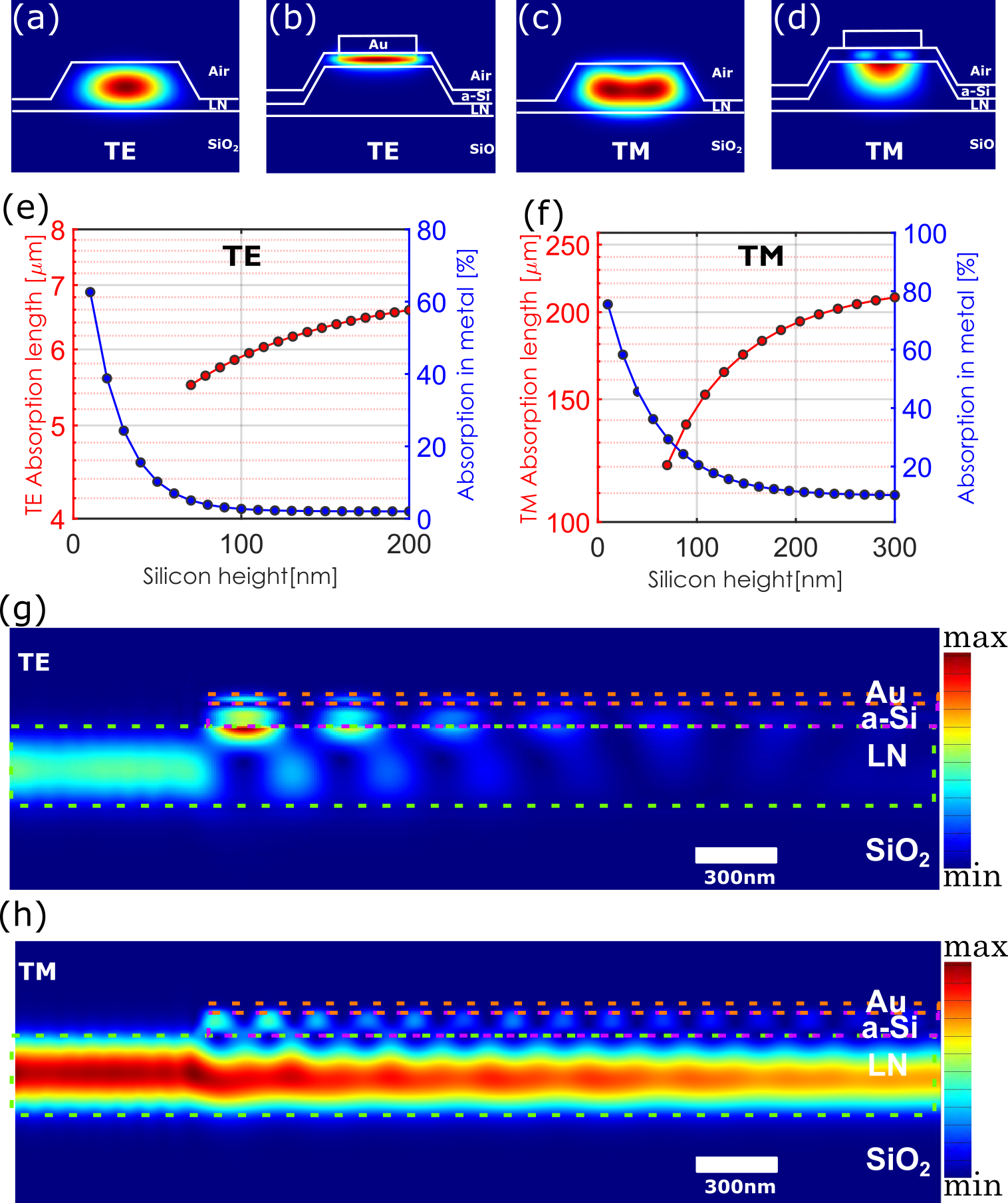}
	\caption{\label{fig2}\textbf{(a-d)} Simulated TE and TM optical modes of LN waveguide and integrated photodetector (thickness of a-Si layer-50nm) respectively.  \textbf{(e-f)} Simulated coupling length and total amount of optical absorption in the metal versus thickness of\ the a-Si layer for TE and TM modes respectively.   \textbf{(g-h)} Finite difference time domain simulation of the electric field intensity  \(  \vert E \vert ^{2} \)  in the coupling section for TE and TM polarizations respectively. Green, purple and orange lines represent the boundaries of the LN waveguide, a-Si layer, and gold contact, respectively. 
}

\end{figure}

Our devices were fabricated on LN-on-insulator (LNOI) chips with 300nm X-cut LN layer on top of 2-$ \mu $m thick thermally grown silicon dioxide layer (NanoLN). First, the photonic structures were defined in electron-beam resist by using an electron beam lithography tool (Elionix) and then the pattern was transferred into the LN layer by using Ar+ plasma based reactive ion etching (RIE) tool (etch depth of 250nm). Next, the devices were covered by 1-$ \mu $m silicon dioxide layer by using plasma-enhanced chemical vapor deposition (PECVD) tool (for optical and electrical insulation) and detector areas were opened by using optical photolithography step followed by RIE of silicon dioxide. Next, the absorption layer of p-doped a-Si was deposited on top of the photonic devices using PECVD followed by a rapid thermal processing (RTP). Next, a-Si was patterned and removed from the chip except the active areas of photodetector\ by using additional photolithography step and RIE.  Finally, gold electrodes were formed by using photolithography and metal lift-off process. After the fabrication, the waveguides facets where diced and polished. A false-color Scanning Electron Microscope (SEM image of the fabricated device is shown in Fig. 3. The detectors are 6 $ \mu $m long. The LN waveguide cross-section is 800x300nm and the electrodes are 2 $ \mu $m wide, with 1 $ \mu $m spacing between them.

\begin{figure}
	\centering
	\includegraphics[angle=0,width=0.5\textwidth]{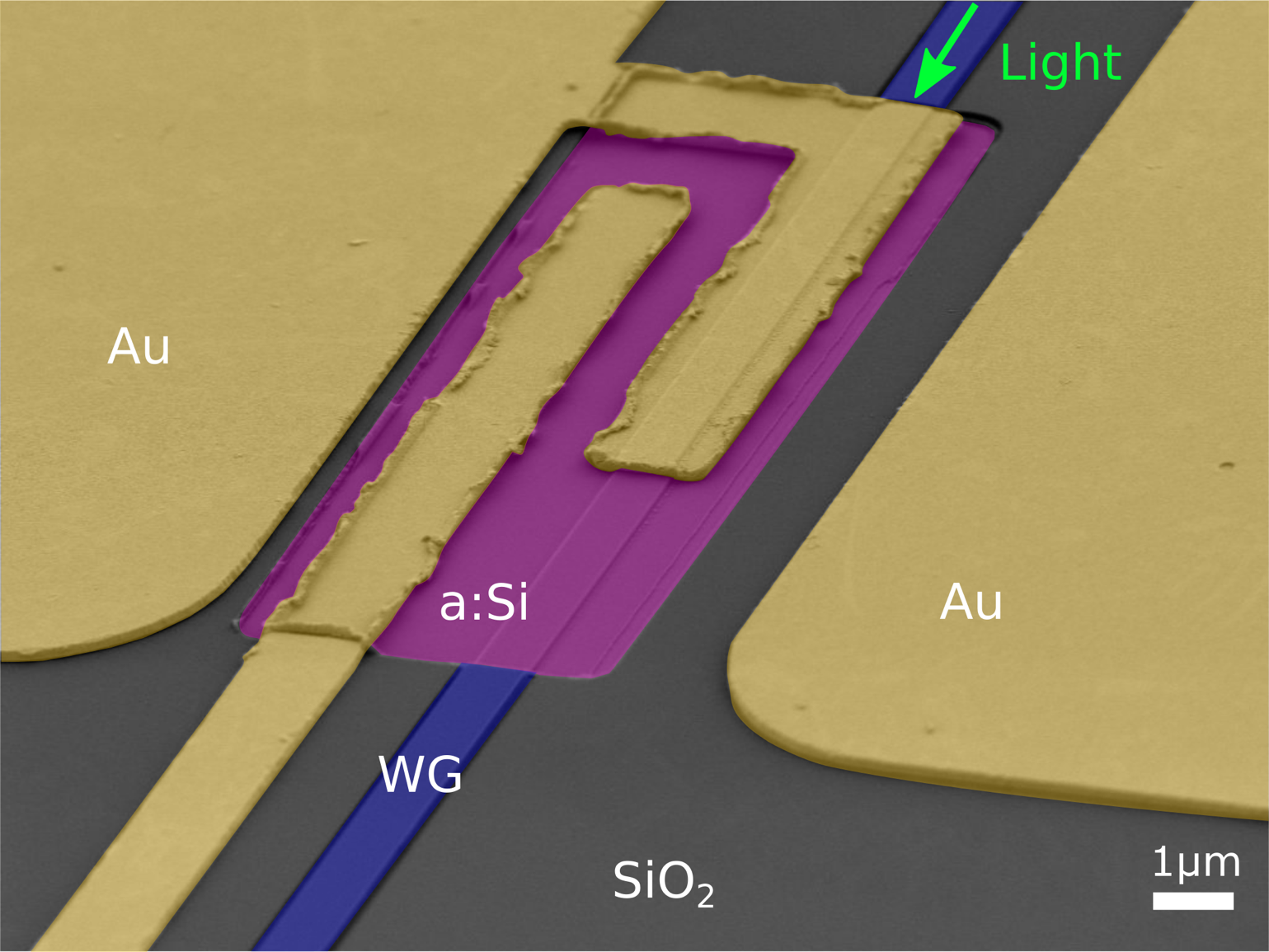}
	\caption{\label{fig3} False colored SEM image of the a-Si layer integrated with LN waveguide (WG). Gold, pink and violet colors represent the metal contacts, a-Si layer and LN waveguide under the silicon dioxide cladding, respectively. 
}
\end{figure}

In order to experimentally characterize the photo-electric response of our detectors, we coupled a TE polarized 635 nm light (from tunable New Focus Velocity laser) into our waveguide. End-fire coupling technique with a lensed fiber with 1$ \mu $m spot size (Oz Optics) was used. Coupling efficiency was estimated to be 10 dB/facet. We note that this can be significantly improved using recently demonstrated spot-size converters[25]. Propagation waveguide loss was estimated to be 0.6 dB/cm by measuring the quality factor of a reference microring resonators fabricated on the same chip (Q $ \sim $  400000). Next, we measured a I-V characteristic of the integrated photodetector (SMU Keithley 2400) for different in-coupled optical powers, as shown in Fig. 4a. The responsivity of the integrated photodetector was found to be 22mA/W at 635 nm wavelength, with dark current of 0.1nA. This value is comparable to previously reported values[18,26] in the case of integrated -Si photodetectors. It should be noticed that while the responsivity of our a-Si detector is about an order of magnitude lower than commercial Si and InGaAs detectors [$\mathbb{R}$(630nm)=0.5  A/W $\mathbb{R}$(630 nm)=0.2  A/W, respectively], our approach offers the important advantages of simplicity and on-chip integration with LN platform. Finally, we measured a responsivity of the photo-detector as a function of a wavelength in 720$-$850 nm range using a M2 SolsTis tunable laser. {Fig.4b  shows an average spectral response of 5 different detectors over this wide wavelength range and the error bars represent the standard deviation from the average.} The maximum responsivity of 37mA/w was achieved at wavelength of 850nm. This is expected since the responsivity of a photodiode (PN, PIN or MSM) typically increases at longer wavelengths according to  \( R= \eta \frac{e}{h \upsilon } \)  where  \( h \upsilon  \)  is the photon energy,  \(  \eta  \)  is the quantum efficiency, and  \( e \)  the elementary charge. At the longer wavelengths (above the bandgap of a-Si) we would expect an abrupt decrease in the responsivity. The exact value of the bandgap of a-Si is strongly dependent on a deposition conditions and techniques and can be varied between 1-3.6eV[27,28]. To estimate the bandgap of our a-Si we have fabricated a reference sample by depositing a 1 $ \mu $m thick film of a-Si layer on top of glass quartz substrate. Next, we measured the transmittance spectrum of the a-Si film by Agilent Cary 60 UV/VIS spectrophotometer. Finally, using a Tauc plot[29] we estimate the bandgap of our a-Si to be 1.4eV.  The responsivity of our photodetector is also strongly dependent on the quality of the a-Si layer. Material defects such as morphology imperfection, voids and grain boundaries can causes to unwanted scattering of the light that will decrease the optical absorption and decrease the responsivity.

\begin{figure}
	\centering
	\includegraphics[angle=0,width=0.5\textwidth]{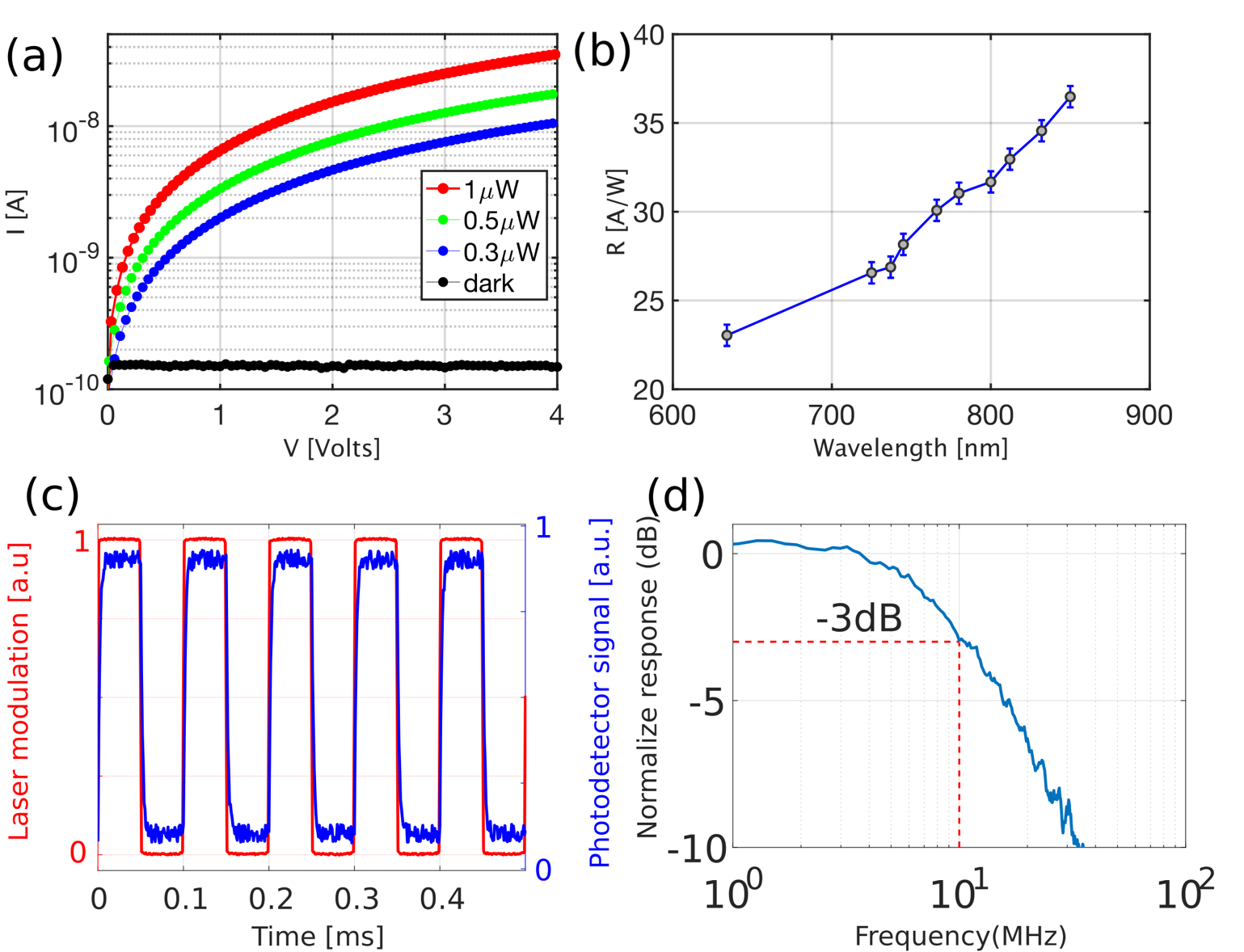}
	\caption{\label{fig4} Experimental characterization of a-Si MSM photodetector.  \textbf{(a)} IV response measured at wavelengths of 635 nm for different optical powers inside the waveguide.  \textbf{(b)} Responsivity as a function of wavelength, at -0.5 V bias.  \textbf{(c)} Temporal response for an input optical square wave signal at 1 KHz.\   \textbf{(d)} Normalized frequency response of integrated photodetector shows a 3dB roll-off frequency at 10Mhz, limited by the performance of trans-impedance amplifier (TIA). The red line indicates the -3dB threshold of TIA.}

\end{figure}

Next, we evaluated the temporal-response of a-Si integrated photodetector by directly modulating the tunable laser using signal at frequency of 10kHz. The detected electrical signal from the photodetector was amplified by trans-impedance amplifier (TIA) (AD8488[30] amplifier with expected cut-off frequency at 10MHz) and displayed on an oscilloscope (Fig.4c). The electrooptical bandwidth of our detector was measured using vector network analyzer (VNA). The laser was modulated by VNA and the detected electrical signal was amplified by TIA. Fig.4d shows a normalized (relative to the coaxial cable loss) frequency response of our photodetector with a 3dB roll of frequency of 10MHz, limited by the cut-of frequency of TIA (shown by red line). While low, this bandwidth is sufficient for several on-chip applications of interest, including power monitoring, frequency-modulation spectroscopy[31], and Pound$–$Drever$–$Hall laser-locking technique[32]. By further optimization of the deposition conditions of the a-Si layer or leveraging commercially available polysilicon deposition foundries, it may be possible to realize a high speed (tens of GHz) optical integrated a-Si photodetector. For example, such high-speed photodetectors have been previously demonstrated with hydrogenated a-Si layer[33,34]. This would in turn enable additional applications in visible light communication[35] and ultrafast optical  characterization[36].

Finally, we integrated our a-Si photodetector with microring resonator device as shown in Fig.5a. In this case, the length of photodetector was chosen to be 5 $ \mu $m in order to allow for only partial absorption of the optical energy, about 50 $\%$  in our case. The remaining optical energy was transmitted through the photodetector and was collected by lensed fiber and detected using the commercial photodetector (PDA36A, Thorlabs, Responsivity 0.65A/W) for comparison measurements. The spectral response of our integrated device was measured using tunable laser, and results are shown in Fig. 5b.

\begin{figure}
	\centering
	\includegraphics[angle=0,width=0.5\textwidth]{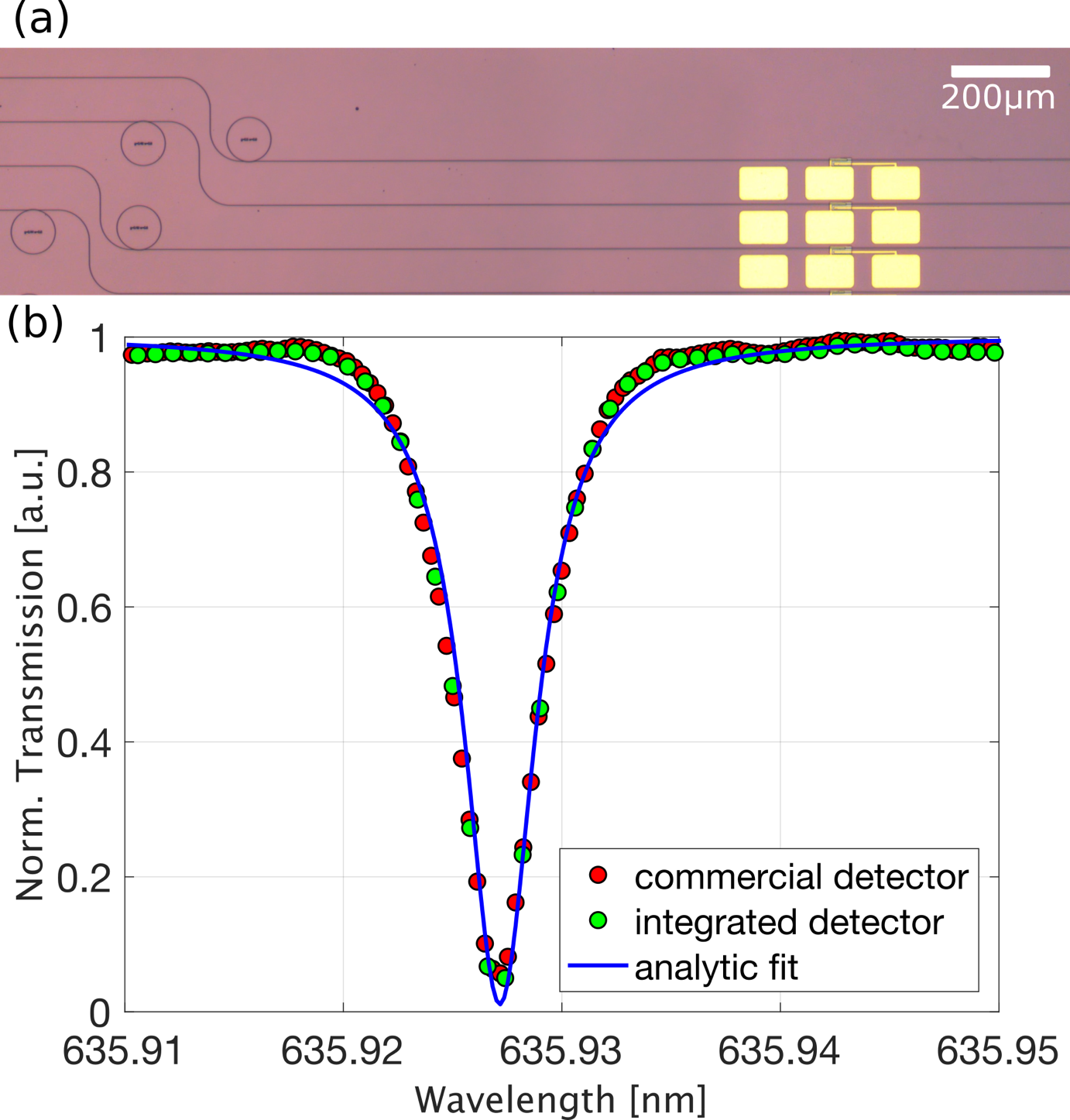}
	\caption{\label{fig5}  \textbf{a)} Optical image of fabricated a-Si photodetector integrated with microring cavity.  \textbf{(b)} Measured transmission spectrum of integrated LN microring resonators. Red and green dots: optical signal measured by integrated a-Si photodetector and external commercial Si photodetector, respectively. Blue line: a fit to Lorentzian function with a loaded Q factor estimated to be  \( 1.5 \times 10^{5} \) 
}
\end{figure}

Fig. 5b shows a transmission spectrum of the fabricated microring resonator measured with both integrated a-Si photodetectors and commercial photodetector. By fitting the experimental result to a Lorenzian function we estimated a loaded Q factor of fabricated microring resonator to be  \( 1.5 \times 10^{5} \) The value of quality factor is lower than previously reported values[7] due to non-optimized waveguide dimensions and additional high temperature (700C) fabrication steps needed to realize the detector[37] including a-Si\ deposition and  rapid thermal processing (RTP). The measured spectral photo-response of integrated photodetector is in good agreement with the results obtained using commercial photodetector. The high spectral selectivity of the integrated\ photonic device  that consist of microring resonator filter and the a-Si integrated photodetector makes it a good candidate for realization a large variety of devices for  detection and sensing application such as integrated spectrometers[38] and bio-sensors[11] .

In conclusion, we have demonstrated an integrated a-Si photodetector in LN photonic platform at visible wavelengths. Responsivity of 37mA/W was measured at wavelengths of 850nm, which is the primary wavelength for multimode fiber optical communication systems based on Vertical-Cavity Surface Emitting Laser (VCSEL). The operating bandwidth of integrated photodetector was measured to be 10MHz, limited by transimpedance amplifier we used. Additionally, we have demonstrated a wavelength selective detection by monolithically integrating photodetector with a microring resonator with quality factor of 150,000.  Furthermore, using a different materials\ (e.g. germanium) for the absorption layer with different energy bandgap it will possible to enhance the bandwidth of the photodetector toward infrared and mid infrared spectral window. Another promising approach for realization a broadband  and high speed integrated photodetectors is to use emerging 2d materials[5,39]. We believe that integrated LN photonic platform will become a promising candidate for realization of multi-element monolithic photonic circuits.

Lithium niobate devices were fabricated in the Center for Nanoscale Systems (CNS) at Harvard, a member of the National Nanotechnology Infrastructure Network, supported by the NSF under award no. 1541959.This work is supported in part by the National Science Foundation (NSF) (ECCS-1740296 E2CDA, IIP-1827720), Defense Advanced Research Projects Agency (DARPA) (W31P4Q-15-1-0013) and Air Force Office of Scientific Research (AFOSR) (MURI: FA9550-12-1-0389).

\section{REFERENCES}
\begin{enumerate}
\item   D. Miller, Nat Phot. 4, 3 (2010).
\item   H. Chen, H. Fu, X. Huang, X. Zhang, T.-H. Yang, J.A. Montes, I. Baranowski, and Y. Zhao, Opt. Express 25, 31758 (2017).
\item   T.-J. Lu, M. Fanto, H. Choi, P. Thomas, J. Steidle, S. Mouradian, W. Kong, D. Zhu, H. Moon, K. Berggren, J. Kim, M. Soltani, S. Preble, and D. Englund, Opt. Express 26, 11147 (2018).
\item   P. Rabiei and P. Gunter, Appl. Phys. Lett. 85, 4603 (2004).
\item   G. Wang, Y. Zhang, C. You, B. Liu, Y. Yang, H. Li, A. Cui, D. Liu, and H. Yan, Infrared Phys. Technol. 88, 149 (2018).
\item   M. He, M. Xu, Y. Ren, J. Jian, Z. Ruan, Y. Xu, S. Gao, S. Sun, X. Wen, L. Zhou, L. Liu, C. Guo, H. Chen, S. Yu, L. Liu, and X. Cai, Nat. Photonics 13, 359 (2019).
\item   B. Desiatov, A. Shams-Ansari, M. Zhang, C. Wang, and M. Lon{\^{c}}ar, Optica 6, 380 (2019).
\item   R. Hamerly, L. Bernstein, A. Sludds, M. Soljacic, and D. Englund, Phys. Rev. X 9, 021032 (2019).
\item   W.K. Chan, A. Yi-Yan, T. Gmitter, L.T. Florez, J.L. Jackel, E. Yablonovitch, R. Bhat, and J.P. Harbison, IEEE Trans. Electron Devices 36, 2627 (1989).
\item   J.P. Hopker, T. Gerrits, A. Lita, S. Krapick, H. Herrmann, R. Ricken, V. Quiring, R. Mirin, S.W. Nam, C. Silberhorn, and T.J. Bartley, APL Photonics 4, 056103 (2019).
\item   Toshihiro Kamei, Brian M. Paegel, James R. Scherer, Alison M. Skelley, and Robert A. Street, and  Richard A. Mathies, (2003).
\item   A. Joskowiak, M.S. Santos, D.M.F. Prazeres, V. Chu, and J.P. Conde, Sensors Actuators B Chem. 156, 662 (2011).
\item   P. Novo, D.M. França Prazeres, V. Chu, and J.P. Conde, Lab Chip 11, 4063 (2011).
\item   A. Samusenko, V.J. Hamedan, G. Pucker, M. Ghulinyan, F. Ficorella, R. Guider, D. Gandolfi, and L. Pavesi, in 2015 XVIII AISEM Annu. Conf. (IEEE, 2015), pp. 1–4.
\item   M. Moridi, S. Tanner, N. Wyrsch, P.-A. Farine, and S. Rohr, Procedia Chem. 1, 1367 (2009).
\item   S.M. Sze and K.K. Ng, Physics of Semiconductor Devices (Wiley-Interscience, 2007).
\item   P.R. Berger, in edited by A.K. Chin, N.K. Dutta, K.J. Linden, and S.C. Wang (International Society for Optics and Photonics, 2001), p. 198.
\item   M.M. Howerton and T.E. Batchman, J. Light. Technol. 6, 1854 (1988).
\item   B. Desiatov, I. Goykhman, N. Mazurski, J. Shappir, J.B.J.B. Khurgin, and U. Levy, Optica 2, 335 (2015).
\item   W. Li and J.G. Valentine, Nanophotonics 6, 177 (2017).
\item   P.B. Johnson and R.W. Christy, Phys. Rev. B 6, 4370 (1972).
\item   H. Zimmermann, Integrated Silicon Optoelectronics (Springer Berlin Heidelberg, Berlin, Heidelberg, 2000).
\item   R.I. Devlen, E.A. Schiff, J. Tauc, and S. Guha, MRS Proc. 149, 107 (1989).
\item   A.M. Johnson, A.M. Glass, D.H. Olson, W.M. Simpson, and J.P. Harbison, Appl. Phys. Lett. 44, 450 (1984).
\item   A. Shams-Ansari, C. Wang, L. He, M. Lon{\^{c}}ar, M. Zhang, and R. Zhu, Opt. Lett. Vol. 44, Issue 9, Pp. 2314-2317 44, 2314 (2019).
\item   A. Samusenko, V.J. Hamedan, G. Pucker, M. Ghulinyan, F. Ficorella, R. Guider, D. Gandolfi, and L. Pavesi, in 2015 XVIII AISEM Annu. Conf. (IEEE, 2015), pp. 1–4.
\item   R.H. Klazes, M.H.L.M. van den Broek, J. Bezemer, and S. Radelaar, Philos. Mag. B 45, 377 (1982).
\item   Jasruddin, W.W. Wenas, T. Winata, and M. Barmawi, in ICSE 2000. 2000 IEEE Int. Conf. Semicond. Electron. Proc. (Cat. No.00EX425) (IEEE, n.d.), pp. 245–248.
\item   J. Tauc, Mater. Res. Bull. 3, 37 (1968).
\item   Z.-C. Zeng and Z.D. Schultz, Rev. Sci. Instrum. 89, 083105 (2018).
\item   G.C. Bjorklund, M.D. Levenson, W. Lenth, and C. Ortiz, Appl. Phys. B Photophysics Laser Chem. 32, 145 (1983).
\item   R.W.P. Drever, J.L. Hall, F. V. Kowalski, J. Hough, G.M. Ford, A.J. Munley, and H. Ward, Appl. Phys. B Photophysics Laser Chem. 31, 97 (1983).
\item   D.R. Larson and R.J. Phelan, Jr., in edited by M.A. Mentzer (1988), p. 59.
\item   D.H. Auston, P. Lavallard, N. Sol, and D. Kaplan, Appl. Phys. Lett 36, 66 (1980).
\item   T.-C. Wu, Y.-C. Chi, H.-Y. Wang, C.-T. Tsai, Y.-F. Huang, and G.-R. Lin, Sci. Rep. 7, 11 (2017).
\item   D.H. Auston, Appl. Phys. Lett. 26, 101 (1975).
\item   D. Sugak, I. Syvorotka, O. Buryy, U. Yachnevych, I. Solskii, N. Martynyuk, Y. Suhak, G. Singh, V. Janyani, and S. Ubizskii, IOP Conf. Ser. Mater. Sci. Eng. 169, 012019 (2017).
\item   Xiao Ma, Mingyu Li, and Jian-Jun He, IEEE Photonics J. 5, 6600807 (2013).
\item   G. Konstantatos, Nat. Commun. 9, 5266 (2018).
\end{enumerate}

\end{document}